\documentclass[aps,prl,showpacs,twocolumn,groupedaddress,amssymb]{revtex4}
\usepackage{graphicx}% Include figure files
\usepackage{dcolumn}% Align table columns on decimal point
\usepackage{bm}% bold math

\begin{document}

%\preprint{1st paper}

\title{X-ray Raman compression via two-stream instability in dense plasmas}% \\

\author{S. Son}
\email{seunghyeonson@gmail.com}
\affiliation{18 Caleb Lane, Princeton, NJ 08540}
\author{Sung Joon Moon\footnote{Current Address: 28 Benjamin Rush Ln. Princeton, NJ 08540}}
\affiliation{Program in Applied and Computational Mathematics, Princeton University, Princeton, NJ 08544}
\date{\today}% It is always \today, today,

\begin{abstract}
A Raman compression scheme suitable for x-rays, where the Langmuir wave is created by an intense 
beam rather than the pondermotive potential between the seed and pump pulses, is proposed.
%On the contrary to the conventional Raman compression scheme where  an intense Langmuire necessary for the compression is generated by the beat-wave of the pump and seed pulse, the new scheme enhances the Langmuir wave by extracting free-energy from a beam. 
The required intensity of the seed and pump pulses enabling the compression could be mitigated by
more than a factor of 100, compared to conventionally available other Raman compression schemes.
The relevant wavelength of x-rays ranges from $1$ to $10 \ \mathrm{nm}$.
\end{abstract}


\maketitle

The Backward Raman scattering (BRS) is a three wave interaction where the beating
pondermotive potential from the two light waves (the pump and the seed pulses) creates
a Langmuir wave, which in turn channels the energy from the pump to the seed pulse.
Recently, the BRS-based compression schemes for x-ray lasers have been 
studied~\cite{fast2,Free,Fisch,Fisch2,Fisch3,sonlandau,sonpre}. 
One major practical difficulty of this scheme is the requirement of high intensity of
the pump and the seed pulses.
In the visible light regime, the chirped pulse amplification (CPA)~\cite{french}
can be used to overcome this issue.
However, it remains as a challenge for x-rays, where the amplification scheme is
unavailable~\cite{sonband}.
Any progress leading to the mitigation of this requirement would enhance the feasibility
of the x-ray Raman compression. 
%Therefore,  it would critically improve the prospect of the x-ray Raman compression  if there is a way to reduce the intensity requirement. 

The two-stream instability, which has been used in an amplification scheme 
for the low-frequency light waves~\cite{wavetube1,wavetube2, wavetube3},
is a good candidate for this progress.   
Recently, the proton and the electron beams have been generated by irradiating
an intense laser to a thin metal target~\cite{ebeam, beam1, beam2, beam3},
with a view to generate a hot spot in the inertial confinement
fusion~\cite{tabak,lindl, beam4, beam5}.
%These intense proton or electron beams are  crucial for generating the hot spot in the inertial confinement fusion~\cite{tabak,lindl, beam4, beam5}.   
Such beams are so intense that it can make a dense plasma unstable. 
The goal of this paper is to propose a BRS-based x-ray compression scheme utilizing
the two-stream instability of the beam in dense plasmas.

Imagine an intense plasma beam propagating through a background dense plasma.
When the Langmuir wave gets amplified by extracting the free energy available
from the beam (via the two-stream instability), rather than from the beating pondermotive
potential, the resulting Langmuir wave would be much more intense than the one
that would have been excited in a stable plasma with the same seed and pump pulses. 
%Then  the Raman compression might be feasible with the weak pump and seed. 
It will be shown that the seed and the pump pulses, orders of magnitude less intense than
what are needed for the conventional BRS-based x-ray compression~\cite{sonpre, Fisch3},
could be still compressed.
In this case, the pondermotive potential acts as an initial seed
rather than the driver in generating an intense plasmon. 
The relevant background electron density turns out to be $10^{23} \sim 10^{25} \mathrm{cm^{-3}}$ 
and the beam electron density is $10^{22} \sim 10^{24} \ \mathrm{cm^{-3}}$,
where the electron drift energy is of a few $ \mathrm{keV}$.
%which might be available from the proton beam rather than the electron beam. 
The scheme to be proposed here is relevant to the x-rays of the wavelength
$1$ to $10 \ \mathrm{nm}$.
%The compression scenario will be illustrated via simple one-dimensional (1-D)
%fluid equations describing the three-wave interaction.
When considering the two-stream instability for the x-ray BRS,
it should be noted that the physical processes in dense plasmas are different from
those in classical plasmas~\cite{sonpre2,sonpla, sonprl, sonlandau}.  
In particular, the Umklapp process~\cite{sonpre,sonpre2} and
the quantum-mechanical effect~\cite{sonlandau} have been examined.

\begin{figure}
\scalebox{0.5}{
\includegraphics{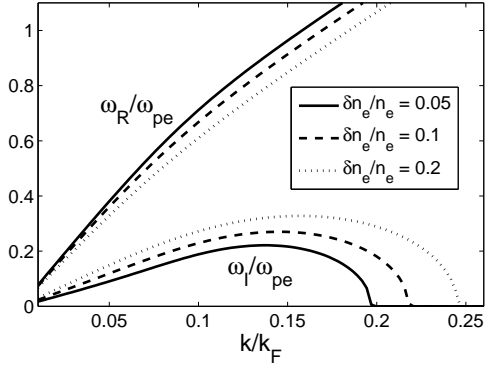}}% Here is how to import EPS art
\caption{\label{fig:1}
The real ($\omega_{R}/ \omega_{\mathrm{pe}}$) and the imaginary
($\omega_I/\omega_{\mathrm{pe}}$) part of the most unstable mode,
%(satisfying the equality $\epsilon = 0$,
as a function of $k/k_F$,   
where $k_F = (3 \pi^2 n_e)^{1/3} $ is the Fermi wave vector,
$n_e = 10^{24} \ \mathrm{cm^{-3}} $, and $m_e V_{b}^2 = 1.25 \ \mathrm{keV}$,
for three values of $\delta n_e/n_e$ = 0.05, 0.1, and 0.2.
}
\end{figure}

Consider an electron beam of the density $\delta n_e$ drifting with the velocity
$\mathbf{V_b}$ through a background plasma of the electron density $n_e$.
The electron beam may have been generated either as an accompanying electron cloud of
an intense proton beam or by an intense laser impinging to a thin metal. 
Let us assume that $|\mathbf{V_b}|$ is much larger than the thermal speed of the
background (beam) electron
$v_{\mathrm{te}} = \sqrt{T_e/m_e}$ ($v_{\mathrm{teb}} = \sqrt{T_{eb}/m_e}$), 
where $T_e$ ($T_{eb}$) is the background (beam) electron temperature.
The dielectric function of the electron can be shown to be
\begin{equation} 
\epsilon(\mathrm{k},\omega) = 1 - \frac{\omega_{\mathrm{pe}}^2 }{ \omega^2}
- \frac{ \delta \omega_{\mathrm{pe}}^2}{ (\omega- \mathbf{k} \cdot \mathbf{V}_{b})^2} \mathrm{,}
\end{equation}
where $ \omega_{\mathrm{pe}}  = \sqrt{4 \pi n_e e^2/m_e}$ and
$\delta  \omega_{\mathrm{pe}}  = \sqrt{4 \pi \delta n_e e^2/m_e}$.   
%For a given $\mathbf{k}$, there are four $\omega_s$'s  satisfying the equation $\epsilon =0$. 
The two-stream instability exists when at least one of the solutions has a positive
imaginary part, i.e., $\omega_I > 0$, where $\omega_s = \omega_R + i\omega_I$ is the most unstable
solution to the relation $\epsilon = 0$.
%Let us define  $ \omega_R =  \mathrm{Re} (\omega_s) $ and 
%$ \omega_I =  \mathrm{Re} (\omega_s) $, where $ \omega_s $ is the solution of the most unstable mode.
In the regime of our interest, $\delta n_e $ is much smaller than $n_e$.
The existence of the two-stream instability in such cases is calculated for a range
of $|\mathbf{k}|$ (Fig.~\ref{fig:1}).

For the seed and the pump pulses in the x-ray Raman compression, a pondermotive 
potential $e\phi_0 \cos(\mathrm{k}\cdot \mathrm{x} - \omega_p t )$ would generate
a Langmuir wave. 
%Now, consider a pondermotive potential 
%$e\phi_0 \cos(\mathrm{k}\cdot \mathrm{x} - \omega_p t )$ 
%from the beat of the pump and the seed in the plasma mentioned above.
According to the linear theory, the Langmuir wave amplitude $ \tilde{n} / n_e $
in a stable plasma has  the maximum of 
\begin{equation}
\frac{\tilde{n}_{\mathrm{max}}}{n_e} =  
\left(\frac{2k^2}{\omega_{\mathrm{pe}} \gamma} \right)
 \left(\frac{e\phi_0}{m_e}\right) \mathrm{,} \label{eq:max}
\end{equation}
where $\tilde{n}$ is the density perturbation from the wave
and $\gamma $ is the plasma damping rate~\cite{sonpre}.
Note that the damping rate $\gamma$ is at least $\omega_{\mathrm{pe}}/100$
due to the Umklapp process in dense plasmas of our interest~\cite{sonpre2}.   
In the presence of the two-stream electron beams, $\tilde{n}_{\mathrm{max}}$
could exceed the value given in Eq.~(\ref{eq:max}). 
If $\omega_p = \omega_R$, the Langmuir wave of the unstable mode would grow 
without any phase lag with respect to the pondermotive potential. 
When $\omega_p \neq \omega_R$, the unstable mode still grows from the instability,
but, as a consequence of the phase oscillation between the Langmuir wave and
the potential,  
%no net energy flows from the pump to the seed pulse.
the direction of the energy flow oscillates between the pump and the seed pulses with 
the oscillation frequency $\sim \ |\omega_p - \omega_R|$ instead of flowing from the pump to the seed. 
Therefore, it is better to match the frequency of the pondermotive potential
to $\omega_R$.

Let us verify the above statement. 
Consider the physical parameters in Fig.~\ref{fig:1} for the case of
$\delta n_e/n_e = 0.1$.
% by a simple 1-D fluid simulation 
%of the two electron beams in the presence of a small pondermotive potential. 
For $k = 0.2 k_F$,  
we integrate the 1-D fluid equations for the two electron beams in the presence of
a small pondermotive potential. 
The continuity, the momentum balance, and the Poisson equations are consistently
integrated using a pseudo-spectral method, in the presence of a potential
$\phi_{\mathrm{pod}} = \phi_0 \cos( k x -\omega_p t)$, 
where $e\phi_0$ is as small as $0.01 \ \mathrm{eV}$. 
The numerical integration shows that the Langmuir wave grows 
with the rate $\omega_I$ and the amplitude eventually exceeds the
maximum amplitude from the linear theory in Eq.~(\ref{eq:max}). 
In particular, when $\omega_p = \omega_R$ is satisfied,
the wave grows with the rate $\omega_I$, while remaining synchronized with
the pondermotive potential.
When $\tilde{n} / n_e =0.4$, the wave does not grow any longer,
rather it starts to break.   
Qualitatively the same behavior is observed in various other parameter regimes not shown
in Fig.~\ref{fig:1}.  

\begin{figure}
\scalebox{0.9}{
\includegraphics[width=0.935\columnwidth]{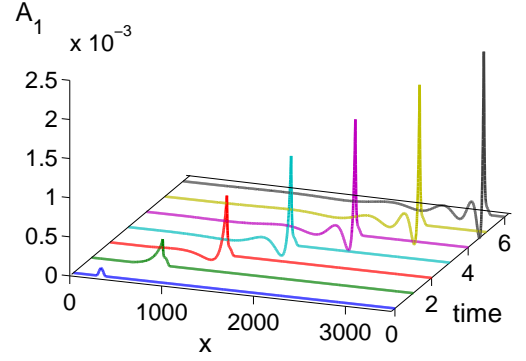}}
\caption{\label{fig:2}
1-D simulation of the Raman compression,
%Time series of the pump pulse $A_1 = eE_1/m_e\omega_1 c$,
%and the Langmuir wave, $ A_3 = \tilde{n} / \bar{n} $,
%traveling to the right through AL (gray area, $ 0\le x \le2000$),
%where $x$ is normalized by $ \omega_{\mathrm{pe}}/c $.
%Each pulse is separated by 5000/$\omega_{\mathrm{pe}}$, and
%the duration of the pump pulse is $4000/\omega_{\mathrm{pe}}$ in both cases.  
%ADD'TL CAPTION:
where $n_e = 10^{24} \ \mathrm{cm^{-3}} $, $\delta n_e = 0.1 n_e$ 
$m_e V_b^2 = 1.25 \ \mathrm{keV}$, $k=0.2 k_F = 6 \times 10^7 \ \mathrm{cm}^{-1}$ 
and $k_F= 3 \times 10^{-8} \ \mathrm{cm}^{-1}$.  
Each pulse is separated by 500/$\omega_{\mathrm{pe}}$ in time,
and $x$ is normalized by $ \omega_{\mathrm{pe}}/c $.
The relevant wavelength of the seed and the pump pulses is 
$\lambda = 2\pi /(k/2) \sim 2\ \mathrm{nm}$
and $\omega/\omega_{\mathrm{pe}} = 15$.  
$\nu_3 = -0.2 \ \omega_{\mathrm{pe}}$ as been computed in Fig.~\ref{fig:1} 
and we neglect the Landau damping and the Umklapp process
($\gamma \sim 0.04 \ \omega_{\mathrm{pe}}$). 
In the simulation, the intensity of the seed and the pump pulses
is $2.5 \times 10^{15} \ \mathrm{W} / \mathrm{cm^2}$. 
}
\end{figure}

The creation of the plasmon with moderate pump and seed pulses has
an important implication on the BRS-based compression of the x-rays.
The 1-D BRS three-wave interaction can be described by~\cite{McKinstrie}:
\begin{eqnarray}
\left( \frac{\partial }{\partial t} + v_1 \frac{\partial}{\partial x} + \nu_1\right)A_1  = -ic_1 A_2 A_3  \nonumber \mathrm{,}\\
\left( \frac{\partial }{\partial t} + v_2 \frac{\partial}{\partial x} + \nu_2\right)A_2  = -ic_2 A_1 A^*_3   \label{eq:2} \mathrm{,} \\
\left( \frac{\partial }{\partial t} + v_3 \frac{\partial}{\partial x} + \nu_3\right)A_3  = -ic_3 A_1 A^*_2  
\nonumber \mathrm{,}
\end{eqnarray}
where $A_i= eE_i/m_e\omega_ic$ is 
the ratio of  the electron quiver velocity of the pump pulse ($i=1$)
and seed pulse ($i=2$), relative to the velocity of the light $c$, 
and $A_3 = \tilde{n_e}/n_e$ is the the Langmuir wave amplitude.
$\nu_1 $ ($\nu_2$) is the rate of the inverse bremsstrahlung 
of the pump (seed), $\nu_3$ is the plasmon decay rate, 
$ c_i = \omega_{pe}^2/2\omega_i$ for $i=1$ and 2, $c_3 = (cq)^2/2\omega_3$,
and $\omega_{1} (\omega_{2})$  is the frequency of the pump (seed) pulse.
In the regime of our interest, $\nu_1 $ ($\nu_2$) is small 
and almost negligible as shown by Son, Ku and Moon~\cite{sonpre}. 
%For details, see ~\cite{sonpre}. 
In the absence of the two-stream instability, $\nu_3$ becomes positive
due to the Landau damping and the Umklapp process. 
When the two-stream instability dominates over the damping, 
$\nu_3$ becomes negative ($\nu_3 \cong \omega_I$). 
We also assume that, when $A_3 > 0.4$, $\nu_3 $ is positive and large
due to the wave breaking. 
A simple 1-D simulation of the above equation based on the pseudo-spectral methods
shows that the pump pulse can be compressed by a factor of, as big as 400 (Fig. \ref{fig:2}).  
It should noted that the intensity used in this numerical calculation
($2.5 \times 10^{15} \ \mathrm{W} / \mathrm{cm^2}$) is less than the
one used in Ref.~\cite{sonpre}, which is the minimum intensity of the seed
and the pump pulses for the x-ray compression in a stable plasma,
by more than a factor 100.
%Given the fact that 
%the intensity of the pump and the seed is less than by a factor 100,
%this  suggests that the intensity requirement of the pump and seed 
%can be reduced significantly so that a small facility with 
%moderate x-ray laser might compress the x-ray to a duration of 
%femto-seconds. 

To summarize, a new scheme of the backward Raman compression is proposed.
This scheme would enable the compression of x-rays, using a pump and seed pulse of
as little as 0.01 times what is required for the case of stable plasmas. 
However, this scheme requires a powerful beam which, in turns,
needs a powerful visible light laser of at least kJ power.
This trade-off is not a concern thanks to currently available powerful lasers
in the visible light regime.
% with the current technology limit being overcome in fast phase, as evident in the inertial confinement fusion and other intense laser technologies.
When this scheme becomes realized,
a small x-ray laser facility with moderate laser power would enable the x-ray Raman compression. 

One possible concern would be the premature Backward Raman pump depletion 
from the instability of the background noise plasmons, which
can be overcome by the reduced Landau damping of the x-ray suppressed
via the band gap effect~\cite{sonlandau}.   
Consider the physical parameters used for the example in Fig.~\ref{fig:2},
and let us assume that the electron temperature of the background plasma 
is $400 \ \mathrm{eV}$. 
Then the damping contribution from the Umklapp process~\cite{sonpre2} is $0.02
\ \omega_{\mathrm{pe}}$ and the Landau damping is $ 0.2  \ \omega_{\mathrm{pe}}$. 
Most instabilities of noise plasmons  would be suppressed 
by the Umklapp process and the Landau damping. 
However, the Landau damping from the pondermotive potential 
would be reduced  considerably due to the band gap effect~\cite{sonlandau}, 
where the classical bounce frequency is much lower than $ \hbar k^2 / 2 m_e$
and the amplification of the Langmuir wave is still possible.  
%In order for self-consistency, the above statement should be verified by 
%the quantum particle-in-cell simulation.
% using the similar approach as the
%Hartree-Fock approximation, 
%which will take into accounts  the Landau damping suppression, 
%two-stream instability and the wave breaking of the Langmuir wave. 
%We leave this to the future research.  
More detailed estimation should be addressed through the quantum particle-in-cell simulation using the mean-field theory (the so-called Hartree-Fock approach), which is in progress.

The drift energy of the accompanying electrons of the proton beam is
much lower (a few keV) than the that of the electron beam directly generated
from the laser-metal interaction (10 MeV). 
It is well-known that there is no instability if $k V_b \gg \omega_{\mathrm{pe}} $.
As a high-k and high frequency X-ray is preferred, the proton beam is preferred to 
the electron beam in the Raman compression.

\bibliography{two}% Produces the bibliography via BibTeX.

\end{document}